\documentclass[a4paper,11pt]{article}
\usepackage[T1]{fontenc}
\usepackage{newtxtext,newtxmath}
\usepackage[scaled=0.9]{inconsolata}
\usepackage{amsmath,bm}
\usepackage{graphicx}
\usepackage{booktabs,tabularx,array}
\usepackage[margin=25.4mm]{geometry}
\usepackage{enumitem}
\usepackage{float}
\usepackage{placeins}
\usepackage{url}
\usepackage{microtype}
\usepackage{caption}
\usepackage{titlesec}
\usepackage{balance}
\usepackage[hidelinks]{hyperref}
\hypersetup{pdfauthor={Akito Hattori},pdftitle={LINE Conversation History Retrieval for Personal Memory RAG}}
\setlist{nosep}
\titleformat{\section}{\large\bfseries}{\thesection.}{0.55em}{}
\titleformat{\subsection}{\normalsize\bfseries}{\thesubsection}{0.55em}{}
\titleformat{\paragraph}[runin]{\normalsize\bfseries}{}{0pt}{}[.]
\titlespacing*{\section}{0pt}{1.4em}{0.55em}
\titlespacing*{\subsection}{0pt}{1.0em}{0.35em}

\newcommand{\code}[1]{\texttt{\detokenize{#1}}}

\title{\textbf{LINE Conversation History Retrieval for Personal Memory RAG:\\Evaluating Search Representations and Hybrid Retrieval}}

\author{Akito Hattori\\\normalsize Independent Researcher, Tokyo, Japan}
\date{August 2026}

\begin{document}
\pagestyle{plain}
\twocolumn[{
\begin{@twocolumnfalse}
\begin{center}
{\LARGE\bfseries LINE Conversation History Retrieval for Personal Memory RAG:\\Evaluating Search Representations and Hybrid Retrieval\par}
\vspace{1em}
{\normalsize Akito Hattori\\Independent Researcher, Tokyo, Japan\par}
\vspace{0.55em}
{\normalsize August 2026\par}
\vspace{0.8em}
\end{center}

\begin{center}
\begin{minipage}{0.92\textwidth}
\begin{center}\textbf{Abstract}\end{center}
\small
As an initial step toward personal memory retrieval-augmented generation (RAG) for large language models (LLMs), this study presents a retrieval-only case study over one user's LINE conversation history. We segmented 358,896 messages into 22,329 temporally coherent chunks and constructed three search representations: \path{raw_text}, a generated summary, and \path{embedding_text}, which combines a summary with a raw-text excerpt and other fixed text. We compared BM25, dense vector retrieval, and linear hybrid retrieval on 100 evaluation questions verified by a single annotator. Among individual retrievers, \path{embedding_text_bm25} achieved the highest point estimate, with Recall@5 of 0.584. We then explored six retriever pairings and 21 weights, for 126 configurations on the same evaluation set. The selected combination of \path{embedding_text_bm25} and \path{embedding_text_vector} at $\beta=0.45$ achieved Recall@5=0.697, MRR@5=0.595, and nDCG@5=0.575. Its Recall@5 exceeded that of \path{embedding_text_bm25} by 0.113, with a question-level paired percentile-bootstrap 95\% confidence interval of [0.048, 0.184]. This interval is conditional on fixing the configuration selected on the same 100 questions and does not account for uncertainty from configuration selection or weight search. The difference from a summary-based hybrid at $\beta=0.50$ was 0.050, with a 95\% confidence interval of [$-0.013$, 0.115], so no clear difference could be established. The 17 aggregate questions also yielded lower point estimates than the other question types, suggesting that flat chunk-level retrieval struggles when evidence is distributed across multiple times and conversations. This evaluation is an exploratory single-user, single-annotator study conducted on the same question set used for configuration search; it does not evaluate final answer generation or generalization to unseen questions.
\end{minipage}
\end{center}
\vspace{1em}
\end{@twocolumnfalse}
}]
\raggedbottom

\section{Introduction}
\subsection{Background}
Recent large language models (LLMs) have demonstrated strong performance in text generation, summarization, question answering, and reasoning. By default, however, an LLM does not know an individual user's past experiences or conversation history, which limits its ability to make experience-grounded suggestions or answer questions about prior events. Practical personal assistance often depends on user-specific context, including past conversations, commitments, schedules, relationships, and changes in emotion. If such personal memories can be organized as a searchable database, an LLM could potentially act as a personal assistant that retrieves and uses this context. We refer to this direction as Personal Memory Retrieval-Augmented Generation (RAG). Making personal conversation histories retrievable for RAG may provide a foundation for more personalized assistance. Everyday histories from messaging applications such as LINE (LY Corporation, Tokyo, Japan) contain substantial amounts of such personal context. As an initial investigation, we therefore conduct a case study using the LINE history of a single user.

\subsection{Challenges in Retrieving Personal Conversations}
Unlike conventional documents, histories from messaging applications such as LINE contain many short utterances whose meaning depends heavily on surrounding context, including acknowledgments and fragmentary expressions. They also mix typographical errors, abbreviations, stickers, images, and URLs. As a result, the wording of a user's question often fails to match the wording of the relevant messages directly, making retrieval difficult.

Searching the original text can be effective when the query contains the same words used in the conversation, but it may be less effective for abstract questions about the meaning or background of an exchange. We therefore compare the original text, a natural-language summary of the conversation, and a search representation that combines the summary with a raw-text excerpt. Section~\ref{sec:search-representations} describes how these representations were constructed.

\subsection{Objective}
This study investigates retrieval as a foundation for Personal Memory RAG, in which an LLM can consult a personal conversation history. Rather than evaluating a complete RAG system, we focus on whether the system can retrieve conversation chunks that support an answer. The evaluation is therefore retrieval-only and excludes final answer generation by an LLM.

\paragraph{Research Questions}
\begin{description}[leftmargin=3.4em,style=nextline]
\item[RQ1] Within the implemented retriever--representation configurations, how do \path{raw_text}, summary, and \path{embedding_text} perform as search representations?
\item[RQ2] How do BM25, dense retrieval, and their linear hybrid differ in Top-5 retrieval performance over a personal conversation history?
\item[RQ3] For which questions does retrieval fail, and what additional design elements may be needed for Personal Memory RAG?
\end{description}
\subsection{Contributions}
This study makes three contributions:

\begin{enumerate}[leftmargin=2em]
\item We construct a retrieval-only evaluation environment for Personal Memory RAG using 358,896 LINE messages from one user, segmented into 22,329 chunks.
\item We compare BM25, dense retrieval, and hybrid retrieval across implemented configurations using \path{raw_text}, summary, and \path{embedding_text}. \path{embedding_text_bm25} attains the highest point estimate among individual retrievers, while the explored hybrid of \path{embedding_text_bm25} and \path{embedding_text_vector} attains the highest point estimate on the same evaluation set.
\item Through failure analysis, we observe lower performance on aggregate questions in this case study and identify the difficulty of retrieving distributed evidence with flat chunk-level retrieval alone.
\end{enumerate}
\section{Related Work}
\subsection{Retrieval-Augmented Generation}
RAG retrieves external knowledge and adds it to an LLM's input context, enabling access to knowledge not contained in the model, changing information, and instance-specific context\cite{ref1}. The present study does not evaluate answer generation and instead focuses on the retrieval stage that precedes it.

\subsection{Personal Memory / Personal Information Management}
Prior work on personal information management, lifelogging, and personal knowledge has investigated how people can retrieve and recall past activities and records\cite{ref2,ref3,ref4,ref5}. More recent work has proposed mechanisms for maintaining long-term personal memory in LLM-based systems\cite{ref13}. However, comparisons of search representations for short, everyday, private conversation histories such as LINE remain limited. This study treats a personal conversation history as external memory for Personal Memory RAG.

\subsection{Conversational Search / Chat History Retrieval}
Conversation-history retrieval differs from conventional document retrieval because speakers, chronology, and surrounding context are central, while individual chat messages are often short and elliptical\cite{ref6,ref7}. Existing long-term conversational-memory benchmarks also identify the integration of information distributed across sessions as a challenge\cite{ref14,ref15}. We therefore chunk the LINE history into temporally coherent units rather than treating each utterance as an independent retrieval unit.

\subsection{Sparse, Dense, and Hybrid Retrieval}
Sparse retrieval methods such as BM25 are strong at lexical matching, whereas dense retrieval can capture paraphrases and semantic similarity\cite{ref8,ref9}. Hybrid retrieval combines the two. Personal conversation histories require both exact names, observed vocabulary, and colloquial expressions, as well as more abstract meaning\cite{ref10}. We therefore compare BM25, dense retrieval, and hybrid retrieval.

\section{Dataset and Search Representations}
\label{sec:dataset}
\subsection{LINE History Dataset}
We use a single user's LINE history containing 358,896 messages and segment it into 22,329 chunks (Table~\ref{tab:dataset}). Each chunk is a retrieval unit. To limit disclosure of personal information, we do not release raw message text, real names, source-file names, or specific conversation content; all examples in the paper are synthetic.

\begin{table}[!t]
\centering
\caption{Dataset statistics}\label{tab:dataset}
\small
\renewcommand{\arraystretch}{1.08}
\begin{tabularx}{\columnwidth}{@{}lX@{}}
\toprule
Item & Value \\
\midrule
LINE messages & 358,896 \\
Chunks & 22,329 \\
Evaluation questions & 100 \\
Main evaluation unit & chunk \\
Question construction & NotebookLM-assisted, human-verified \\
Gold annotation & Single annotator \\
\bottomrule
\end{tabularx}
\end{table}

\subsection{Chunking}
The input consists of UTF-8 JSON files in which each date is a key and the messages for that date form the corresponding value. For each message, we read the time, sender, type, and text, and preserve its order in the JSON as \path{source_order}. Timestamps were interpreted as Japanese local time as recorded by the LINE export, but the database does not store a time-zone offset. A missing time was marked as unknown and mapped to the date at 00:00:00 plus \path{source_order} seconds; a missing sender was stored as SYSTEM, and a missing type as text. We applied neither deduplication nor blanket exclusion of media or system events, and retained records with empty text.

Each \path{source_file} was treated as an independent conversation source. Within each source, messages were stably sorted in ascending order by the pair \path{occurred_at} and \path{source_order}. We first split sessions wherever the gap between consecutive messages was strictly greater than 180 minutes. Each session was then chunked from the beginning with a sliding window of at most 42 messages and a stride of 36 messages, producing a six-message overlap between adjacent windows. A final window containing fewer than 42 messages was retained, and no window crossed a source or session boundary. If the rendered input for summary generation exceeded 5,000 Japanese characters, messages were removed one at a time from the end of the window until the input fit the limit. This process produced 22,329 chunks from 358,896 messages. Each message ID was derived from a truncated SHA-1 digest of the source name, \path{source_order}, and message content; each chunk ID was derived from a truncated SHA-1 digest of the source name, within-source index, and constituent message-ID sequence.

\subsection{Search Representations}
\label{sec:search-representations}
We prepared three search representations for each chunk (Table~\ref{tab:representations}). The conversation and timestamp below are English renderings of a synthetic Japanese example used only to illustrate the structure; they are not taken from the actual LINE data.

\paragraph{raw\_text}
This transcript-like representation concatenates LINE messages in timestamp--speaker--text format. It preserves the vocabulary, proper names, and colloquial expressions that actually appeared. However, the prevalence of short utterances, backchannels, and omissions can make an isolated chunk difficult to interpret. The mean \path{raw_text} length was 635 Japanese characters, the 95th percentile was 1,571, and the maximum was 10,031.

\begin{quote}\small A: What time tomorrow?\\ B: Around 6 p.m.\\ A: I may be late because I have cram school.\end{quote}
\paragraph{summary}
This representation is an AI-generated natural-language summary of each chunk. Summaries were generated with the OpenAI Responses API using the requested model \path{gpt-4o-mini}. The input consisted of up to the first 7,000 Japanese characters of \path{raw_text}; the prompt requested approximately 200--500 Japanese characters of output. We set \path{max_output_tokens}=700, a client timeout of 90 seconds, and a 0.2-second interval between API calls. Temperature, top-p, seed, and an explicit retry count were not set. The implementation stopped at the first API exception and committed the database after every successful chunk. Appendix~A gives the actual system and user prompt in English translation and identifies the accompanying file containing the original Japanese text. All 22,329 chunks have \path{summary_status} set to openai, but retry counts and response-model snapshots were not recorded; failure history and exact per-chunk model provenance therefore cannot be audited retrospectively. Fifteen chunks (0.067\%) exceeded the 7,000-character input threshold. Summary length averaged 273 Japanese characters, with a 95th percentile of 446. We did not conduct a pre-specified human quality audit of the generated summaries, so rates of unsupported additions and important omissions remain unmeasured.

\begin{quote}\small The speakers discuss a meeting time, and A indicates that a cram-school commitment may cause a delay.\end{quote}
\paragraph{embedding\_text}
The \path{embedding_text} representation was assembled programmatically for each chunk. It concatenated, on separate lines and in this order, a fixed Japanese sentence identifying the text as a search chunk derived from the user's LINE history, the AI-generated summary, the Japanese label corresponding to ``Original-text excerpt:'', and up to the first 4,000 Japanese characters of \path{raw_text}. Thus, the raw excerpt was appended programmatically to the summary API response; the LLM did not generate the complete \path{embedding_text}. Twenty-two chunks (0.099\%) had \path{raw_text} longer than 4,000 characters. A separate function also shortened inputs immediately before calls to the embedding API; its default retained a combined 2,000 characters from the beginning and end, but the runtime argument was not saved. The database-level construction of \path{embedding_text} is reproducible, but the final character span sent for each chunk when the existing vector cache was generated cannot be fully audited.

\begin{quote}\small
This is a natural-language search chunk derived from the user's LINE history.\\
A and B discuss when to meet the next day. B proposes meeting at around 6 p.m., and A says that a cram-school commitment may cause a late arrival.\\
Original-text excerpt:\\
2026-04-10 20:15 A: What time tomorrow?\\
2026-04-10 20:16 B: Around 6 p.m.\\
2026-04-10 20:17 A: I may be late because I have cram school.
\end{quote}
\begin{table*}[!t]
\centering
\caption{Search representations}\label{tab:representations}
\small
\renewcommand{\arraystretch}{1.08}
\begin{tabularx}{\textwidth}{@{}lXXX@{}}
\toprule
Representation & Content & Strength & Limitation \\
\midrule
\texttt{raw\_text} & Transcript-like original messages & Exact words, names, and colloquial terms & Noisy and context-dependent \\
\texttt{summary} & Generated natural-language summary & Abstract semantic cues & May omit lexical details \\
\texttt{embedding\_text} & Fixed prefix, summary, and raw excerpt & Combines semantic and lexical cues & Contains raw-derived text \\
\bottomrule
\end{tabularx}
\end{table*}

\section{Problem Formulation and Evaluation}
\subsection{Retrieval-only Setting}
We evaluate retrieval, rather than a complete Personal Memory RAG pipeline, and therefore do not assess final answer generation by an LLM. The input is a natural-language question, and the output is a ranking of conversation chunks that can support an answer. The retrieval corpus contains 22,329 chunks.

Let $Q$ be the set of questions, $D$ the set of candidate chunks, and
$\operatorname{rel}(q,d)\in\{0,1,2\}$ the relevance of chunk $d$ to question $q$. We define the set of direct answer evidence and the binary-relevant set that also includes supporting evidence as
\begin{align}
G_2(q) &= \{d\in D \mid \operatorname{rel}(q,d)=2\},\\
G_{\geq 1}(q) &= \{d\in D \mid \operatorname{rel}(q,d)\geq 1\}
\end{align}
respectively. For question $q$, let the ordered Top-$k$ results be $\pi_k(q)=(d_1,\ldots,d_k)$ and let their set be $\mathcal{R}_k(q)=\{d_1,\ldots,d_k\}$. We evaluate recovery of the gold sets using both the ranks in $\pi_k(q)$ and membership in $\mathcal{R}_k(q)$.
\subsection{Evaluation Questions}
We supplied NotebookLM (Google LLC) with a JSON export of all LINE messages, each assigned a unique message ID, and used it to generate candidate evaluation questions together with candidate evidence timestamps, speakers, and excerpts. These candidates were not accepted directly as gold labels. We matched the evidence timestamps and excerpts to messages in the local database and recovered candidate chunks through \path{line_chunk_messages}. When an excerpt did not match the proposed chunk, we added candidates through excerpt-based message-level search. One researcher reviewed 108 questions and 910 candidate rows in a local interface and assigned relevance labels of 0, 1, or 2. We retained 100 questions with at least one relevance-2 chunk and excluded the remaining eight because primary gold could not be established. The final 100 questions and their gold labels were fixed before the v3 retriever comparison. Earlier pilots with 10 and 33 questions nevertheless influenced the design of the retrieval methods and evaluation infrastructure, so question construction was not fully independent of method development.

The retained set contains 34 \path{named_entity}, 49 contextual, and 17 aggregate questions. \path{named_entity} questions ask about a specific person, organization, project, event, or object. Contextual questions require surrounding context about matters such as scheduling, requests, consultations, agreements, or a sequence of events. Aggregate questions integrate information across conversations or time points to ask about trends, roles, sustained activities, or emotional changes.

\subsection{Gold Annotation}
We assigned relevance 0 to chunks that could not support an answer, relevance 1 to chunks that supplied supporting information such as a date, person, or background but could not establish the main answer on their own, and relevance 2 to chunks that alone supported the entire answer or an explicit major part of it. We define relevance-2 chunks as primary gold and chunks with relevance $\geq1$ as binary-relevant. The final annotations contain 144 primary-gold and 45 secondary-gold chunks, and every question has at least one primary-gold chunk.

The annotation interface displayed timestamps, speakers, and short evidence excerpts generated by NotebookLM, raw context previews recovered from the local database, and matching scores. Although the v3 judgments did not directly use the evaluated retriever names or their final rankings, candidate recovery was not fully blind. Judgments were limited to the pool of 910 recovered or additionally searched candidate rows, and unjudged chunks outside the pool were treated as relevance 0 during evaluation. If overlapping windows contained the same evidentiary utterance, distinct chunk IDs were counted as separate gold chunks rather than merged into an evidence group. The gold set may therefore be incomplete, and overlap may affect the denominator of Recall. Recall@5 and MRR@5 use primary gold, whereas nDCG@5 uses graded relevance values of 0, 1, and 2.

\subsection{Evaluation Metrics}
We use Top-5 retrieval as the main setting, reflecting a plausible number of candidate chunks to pass to an LLM in Personal Memory RAG. The primary metrics reported are Recall@5, MRR@5, and nDCG@5\cite{ref17}. For failure analysis, we retain the ranking from the same summary-based hybrid and extend only the cutoff to 10 and 20 to diagnose whether primary gold missed at Top 5 is recoverable. We compute 95\% confidence intervals by bootstrapping questions as the resampling unit\cite{ref18}, with $B=5{,}000$ and seed 12,345. Method differences use a paired percentile bootstrap with the same resamples for both methods.

Each metric is calculated per question and then macro-averaged over $|Q|=100$ questions. All questions satisfy $|G_2(q)|>0$. Recall@$k$ is based on primary gold:
\begin{equation}
\operatorname{Recall}@k
=\frac{1}{|Q|}\sum_{q\in Q}
\frac{|\mathcal{R}_k(q)\cap G_2(q)|}{|G_2(q)|}
\end{equation}
Let $r_q=\min\{i\leq k\mid d_i\in G_2(q)\}$, with $r_q=\infty$ if no such result exists. Then
\begin{equation}
\operatorname{MRR}@k
=\frac{1}{|Q|}\sum_{q\in Q}
\begin{cases}
1/r_q, & r_q\leq k,\\
0, & r_q>k
\end{cases}
\end{equation}
For nDCG, we use relevance values 0, 1, and 2 as graded relevance:
\begin{align}
\operatorname{DCG}@k(q)
&=\sum_{i=1}^{k}\frac{2^{\operatorname{rel}(q,d_i)}-1}{\log_2(i+1)},\\
\operatorname{nDCG}@k
&=\frac{1}{|Q|}\sum_{q\in Q}
\frac{\operatorname{DCG}@k(q)}{\operatorname{IDCG}@k(q)}
\end{align}
Here, $\operatorname{IDCG}@k(q)$ is the DCG of the ideal ranking obtained by sorting all judged chunks for that question in descending order of relevance. Unjudged chunks outside the candidate pool are treated as relevance 0. Because every question has primary gold, $\operatorname{IDCG}@5(q)>0$ for all questions.

Let $x_q^{(m)}$ be the value of a metric for method $m$ on question $q$. We define the difference between methods $a$ and $b$ as
\begin{equation}
\widehat{\Delta}_{a,b}=\frac{1}{|Q|}\sum_{q\in Q}\left(x_q^{(a)}-x_q^{(b)}\right)
\end{equation}
In each bootstrap iteration, question IDs are sampled with replacement and the mean difference between the two methods is computed on the same resample. The 95\% interval is given by the 2.5th and 97.5th percentiles of the $B$ differences. However, the best hybrid was selected by maximizing Recall@5 on these same 100 questions over six pairings and 21 weights, for 126 configurations. MRR@5 and nDCG@5 were then computed for that same selected configuration. The bootstrap interval is therefore conditional on fixing a post-selection configuration and does not include uncertainty from configuration selection, weight search, or multiple comparisons.
\section{Retrieval Methods}
\subsection{BM25 Retrieval}
We apply BM25 retrieval\cite{ref11} independently to the \path{raw_text}, summary, and \path{embedding_text} representations.

The BM25 score is
\begin{equation}
\begin{aligned}
\operatorname{BM25}(q,d)
&=\sum_{t\in q}\operatorname{IDF}(t)\\
&\quad\cdot
\frac{f(t,d)(k_1+1)}
{f(t,d)+k_1\left(1-b+b\frac{|d|}{\operatorname{avgdl}}\right)}
\end{aligned}
\end{equation}
and the inverse document frequency is
\begin{equation}
\operatorname{IDF}(t)
=\log\!\left(1+\frac{N-\operatorname{df}(t)+0.5}
{\operatorname{df}(t)+0.5}\right)
\end{equation}
We refer to the resulting retrievers as \path{raw_text_bm25}, \path{summary_bm25}, and \path{embedding_text_bm25}.

The same regular-expression tokenizer was applied to queries and documents. We performed no Unicode normalization. After lowercasing, the tokenizer scanned from left to right, extracting alphanumeric strings, Katakana strings of at least two characters, Kanji strings of at least two characters, and the short terms \code{usb} and \code{pv}. The stopword list contained Japanese demonstratives and generic question or conversation terms corresponding to expressions such as ``this,'' ``that,'' ``what,'' ``how,'' ``about,'' ``summarize,'' ``consult,'' ``talk,'' ``I,'' and ``LINE.'' Non-stopword single-character tokens were discarded except for \code{a} and \code{b}; repeated query tokens contributed repeatedly to the score. Alphanumeric tokens could contain digits, periods, underscores, and hyphens, whereas Hiragana, single-character Kanji, and emoji were generally not extracted. The exact Unicode ranges and Japanese stopword strings are supplied in \path{bm25_tokenizer_spec.txt}. Because we used neither morphological analysis nor character $n$-grams, the results characterize this regex-tokenized BM25 baseline rather than BM25 in general.

We set $k_1=1.2$ and $b=0.75$ and used the natural logarithm for IDF.

Here, $N$ is the total number of chunks in the corpus and $\operatorname{df}(t)$ is the number of chunks containing term $t$. Ties were broken by chunk start time in descending order.

\subsection{Dense Vector Retrieval}
Dense retrieval represents queries and documents in a continuous vector space\cite{ref8,ref9,ref12}. We embedded both documents and queries with the OpenAI Embeddings API model \path{text-embedding-3-small}. API calls did not explicitly set \path{dimensions}, so the default 1,536 dimensions were used; \path{encoding_format="float"} was specified. Each evaluation question was embedded verbatim, without LLM-based query rewriting, an instruction prefix, or Unicode normalization. Chunk vectors were loaded from a precomputed JSON cache, and query vectors from a database cache keyed by question text and model name. All 100 evaluation queries were cache hits. The audited environment used Python 3.12.13, OpenAI Python SDK 2.36.0, and NumPy 2.4.6, but the SDK version and execution date used at embedding-generation time were not stored in the run record.

The cache was loaded into a NumPy \path{float32} matrix. After $\ell_2$-normalizing every row and the query vector, we scanned all 22,329 chunks by matrix--vector inner product. Candidate extraction used \path{argpartition} followed by descending score sort. No external vector database such as Pinecone or ChromaDB was used. We implemented no additional tie-break for exactly equal scores, so such ordering depends on NumPy's candidate-extraction order.

\begin{equation}
s_{\mathrm{vec}}(q,d)
=\cos(\bm e_q,\bm e_d)
=\frac{\bm e_q^{\mathsf T}\bm e_d}
{\|\bm e_q\|_2\,\|\bm e_d\|_2}.
\end{equation}
We did not evaluate \path{raw_text_vector}, which would have required an additional external embedding pass over the complete original text. However, \path{embedding_text_vector} contains excerpts derived from \path{raw_text}, so the study did not avoid external transmission of all raw-derived text. Consequently, this is not a complete representation $\times$ retrieval-method factorial experiment, and RQ1 is interpreted only over the implemented configurations.

\subsection{Hybrid Retrieval}
For each query, we retrieved the Top 200 candidate chunks from BM25 and from vector retrieval. Because their scores have different scales, we independently applied min--max normalization within each retriever's Top-200 list. We then formed the union of both candidate sets and linearly combined the normalized scores.

The parameter $\beta$ is the BM25 weight: larger values favor BM25, whereas smaller values favor vector retrieval. We varied $\beta$ from 0.00 to 1.00 in increments of 0.05 and compared the following pairings:

\begin{itemize}[leftmargin=1.4em,itemsep=0.1em]
\item \path{summary_bm25} + \path{summary_vector}
\item \path{summary_bm25} + \path{embedding_text_vector}
\item \path{raw_text_bm25} + \path{summary_vector}
\item \path{raw_text_bm25} + \path{embedding_text_vector}
\item \path{embedding_text_bm25} + \path{summary_vector}
\item \path{embedding_text_bm25} + \path{embedding_text_vector}
\end{itemize}

Let $P_r(q)$ be the Top-200 set returned for question $q$ by retriever $r\in\{\mathrm{BM25},\mathrm{vec}\}$. The candidate union is
\begin{equation}
P(q)=P_{\mathrm{BM25}}(q)\cup P_{\mathrm{vec}}(q)
\end{equation}
For each retriever, define the minimum and maximum scores within its candidate set as
\begin{align}
m_r(q)&=\min_{d\in P_r(q)}s_r(q,d),\\
M_r(q)&=\max_{d\in P_r(q)}s_r(q,d)
\end{align}
and normalize per question as
\begin{equation}
\widehat{s}_r(q,d)=
\begin{cases}
\dfrac{s_r(q,d)-m_r(q)}{M_r(q)-m_r(q)},
& \substack{d\in P_r(q),\\ M_r(q)>m_r(q)},\\[0.6em]
0, & \text{otherwise}
\end{cases}
\end{equation}
The final hybrid score is
\begin{equation}
\begin{aligned}
s_{\mathrm{hybrid}}(q,d)
&=(1-\beta)\widehat{s}_{\mathrm{vec}}(q,d)\\
&\quad+\beta\widehat{s}_{\mathrm{BM25}}(q,d),\\
&\quad \beta\in\{0.00,0.05,\ldots,1.00\}.
\end{aligned}
\end{equation}
Each retriever contributes its Top 200 results, so the deduplicated union by chunk ID contains at most 400 candidates. If a candidate appears in only one list, its normalized score from the other retriever is set to 0. If the maximum and minimum scores are equal, every normalized score from that retriever is also set to 0. Because the lowest Top-200 score and an absent score both map to 0, $\beta=0$ or 1 does not necessarily reproduce the exact ranking of the corresponding individual retriever. Hybrid-score ties are broken by descending lexicographic order of the SHA-1-derived chunk ID.

To diagnose the ceiling imposed by candidate generation, we computed macro Recall@200 over primary gold. The values were 0.788 for \path{summary_bm25}, 0.827 for \path{summary_vector}, and 0.885 for the oracle union. For the selected \path{embedding_text} pair, the corresponding values were 0.859 for \path{embedding_text_bm25}, 0.888 for \path{embedding_text_vector}, and 0.941 for the oracle union. Some gold chunks are therefore absent after Top-200 candidate generation, limiting what reranking alone can achieve.

\section{Results}
\subsection{Overall Single Retriever Results}
\begin{figure}[!t]
\centering
\includegraphics[width=0.98\columnwidth]{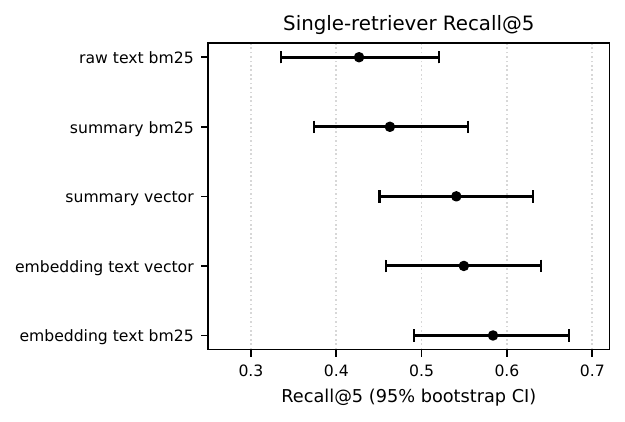}
\caption{Single-retriever Recall@5 with 95\% bootstrap confidence intervals.}\label{fig:single-ci}
\end{figure}

Table~\ref{tab:single} and Figure~\ref{fig:single-ci} report the individual-retriever results. Recall@5 was 0.427 for \path{raw_text_bm25}, 0.463 for \path{summary_bm25}, 0.541 for \path{summary_vector}, 0.549 for \path{embedding_text_vector}, and 0.584 for \path{embedding_text_bm25}. Thus, \path{embedding_text_bm25} had the highest point estimate among individual retrievers. This result is consistent with the hypothesis that placing a summary and raw excerpt together is useful. However, because the fixed description, summary, and raw excerpt were not ablated separately, the contribution of each component cannot be identified.

\begin{table*}[!t]
\centering
\caption{Single retriever performance}\label{tab:single}
\small
\renewcommand{\arraystretch}{1.08}
\begin{tabularx}{\textwidth}{@{}lXXX@{}}
\toprule
Method & Recall@5 (95\% CI) & MRR@5 (95\% CI) & nDCG@5 (95\% CI) \\
\midrule
\texttt{raw\_text\_bm25} & 0.427 [0.335, 0.520] & 0.354 [0.270, 0.439] & 0.348 [0.270, 0.426] \\
\texttt{summary\_bm25} & 0.463 [0.374, 0.555] & 0.419 [0.329, 0.506] & 0.395 [0.313, 0.475] \\
\texttt{summary\_vector} & 0.541 [0.451, 0.631] & 0.445 [0.360, 0.532] & 0.436 [0.359, 0.514] \\
\texttt{embedding\_text\_vector} & 0.549 [0.458, 0.640] & 0.459 [0.375, 0.546] & 0.455 [0.375, 0.536] \\
\texttt{embedding\_text\_bm25} & 0.584 [0.491, 0.673] & 0.491 [0.403, 0.575] & 0.478 [0.399, 0.555] \\
\bottomrule
\end{tabularx}
\end{table*}

The 95\% confidence intervals were computed by percentile bootstrap over evaluation questions ($B=5{,}000$, seed=12,345).

\subsection{Hybrid Retrieval Results}
\begin{figure*}[!t]
\centering
\includegraphics[width=0.78\textwidth]{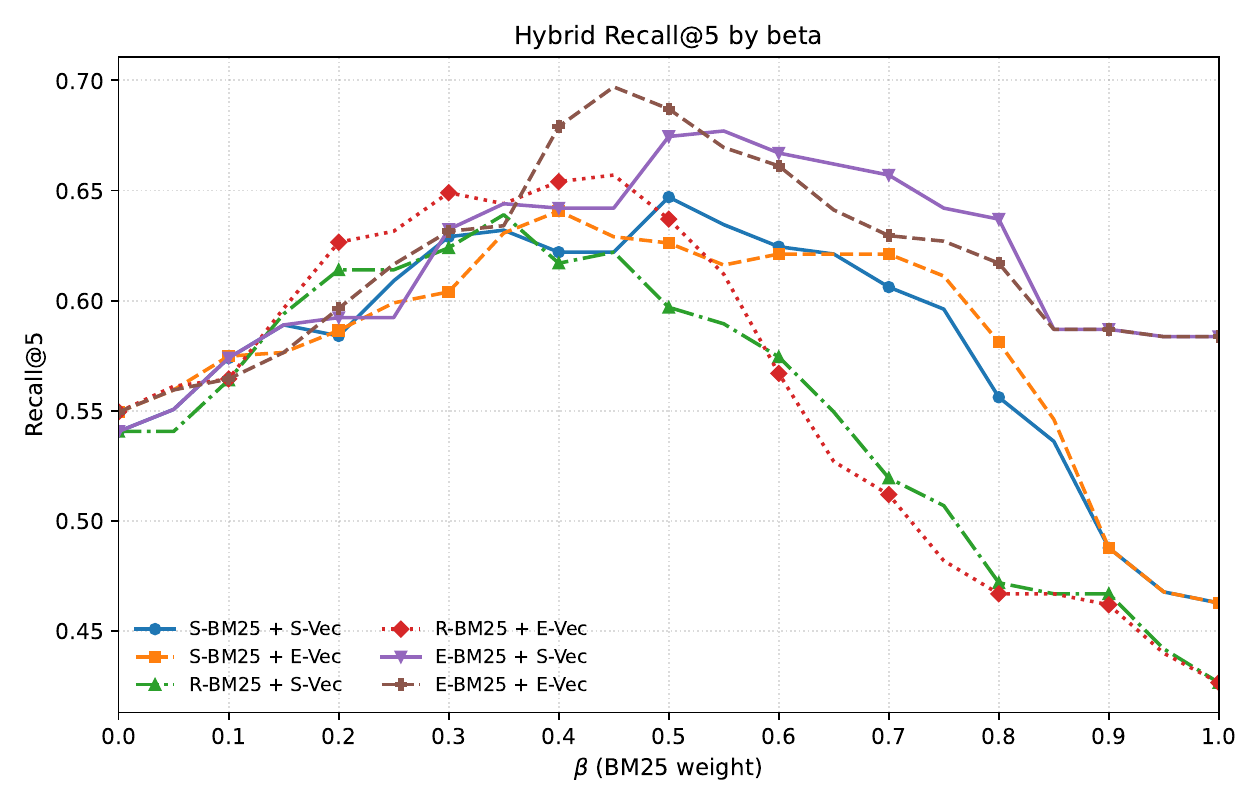}
\caption{Exploratory linear-hybrid Recall@5 versus $\beta$ by retriever combination. R, S, and E denote raw\_text, summary, and embedding\_text.}\label{fig:beta-recall}
\end{figure*}

\begin{figure*}[!t]
\centering
\includegraphics[width=0.78\textwidth]{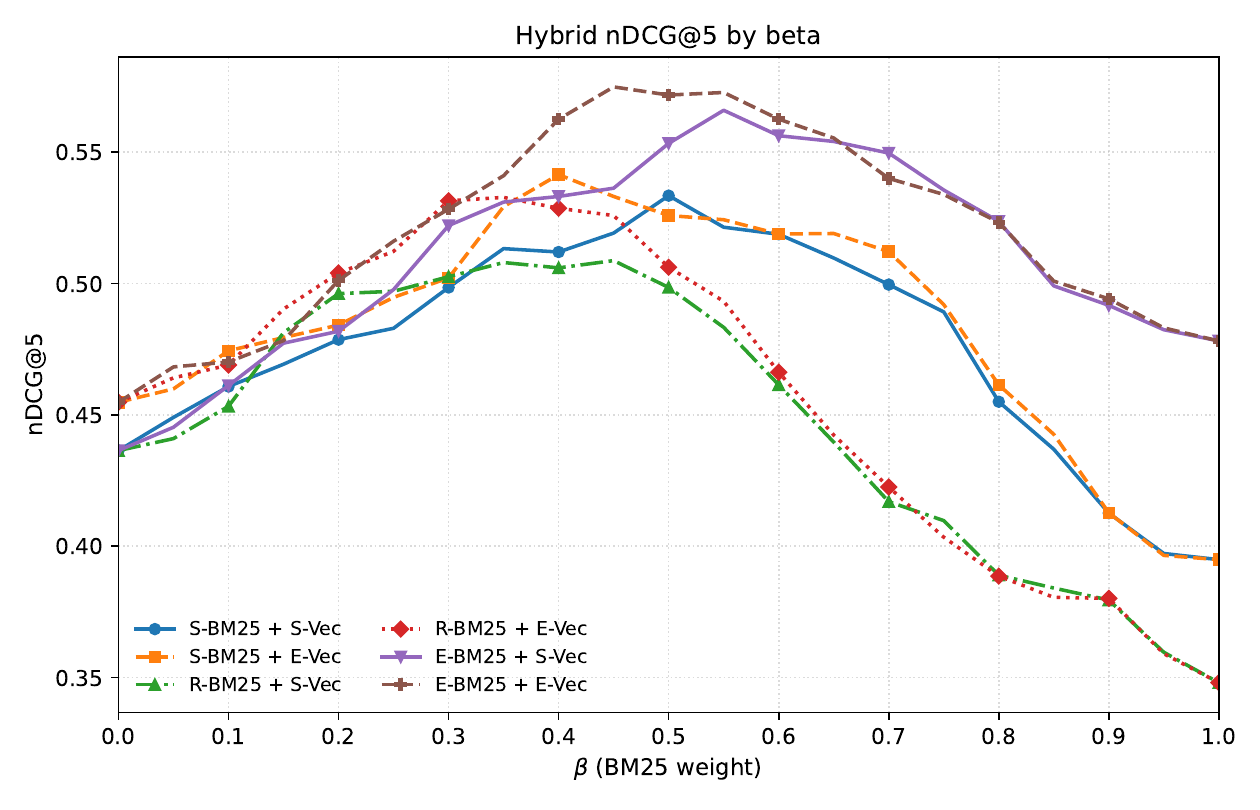}
\caption{Exploratory linear-hybrid nDCG@5 versus $\beta$ by retriever combination. R, S, and E denote raw\_text, summary, and embedding\_text.}\label{fig:beta-ndcg}
\end{figure*}

Hybrid retrieval linearly combined BM25 and dense-retrieval scores. Among the six pairings $\times$ 21 values of $\beta$, for 126 configurations, \path{embedding_text_bm25} + \path{embedding_text_vector} at $\beta=0.45$ produced the highest Recall@5 point estimate. The same configuration achieved Recall@5=0.697, MRR@5=0.595, and nDCG@5=0.575 (Table~\ref{tab:hybrid}). At $\beta=0.50$, its Recall@5 was 0.687. The highest Recall@5 for \path{summary_bm25} + \path{summary_vector} was 0.647 (Figures~\ref{fig:beta-recall} and~\ref{fig:beta-ndcg}). Because the retriever pairing and $\beta$ were selected on the same 100 questions, these are exploratory in-sample comparisons. The \path{embedding_text} hybrid had a higher point estimate than the summary hybrid, but the confidence interval for their difference included 0.

\begin{table*}[!t]
\centering
\caption{Exploratory hybrid comparison selected on the same 100 questions}\label{tab:hybrid}
\small
\renewcommand{\arraystretch}{1.08}
\begin{tabularx}{\textwidth}{@{}lcccc@{}}
\toprule
Hybrid combination & $\beta$ & Recall@5 (95\% CI) & MRR@5 (95\% CI) & nDCG@5 (95\% CI) \\
\midrule
E-BM25 + E-Vec & 0.45 & 0.697 [0.610, 0.775] & 0.595 [0.510, 0.678] & 0.575 [0.496, 0.648] \\
E-BM25 + S-Vec & 0.55 & 0.677 [0.590, 0.758] & 0.590 [0.506, 0.673] & 0.566 [0.487, 0.642] \\
R-BM25 + E-Vec & 0.45 & 0.657 [0.567, 0.740] & 0.528 [0.442, 0.609] & 0.526 [0.448, 0.600] \\
S-BM25 + S-Vec & 0.50 & 0.647 [0.555, 0.735] & 0.548 [0.460, 0.633] & 0.534 [0.453, 0.613] \\
S-BM25 + E-Vec & 0.40 & 0.641 [0.549, 0.727] & 0.556 [0.469, 0.641] & 0.542 [0.460, 0.622] \\
R-BM25 + S-Vec & 0.35 & 0.639 [0.550, 0.725] & 0.522 [0.437, 0.606] & 0.508 [0.432, 0.583] \\
\bottomrule
\end{tabularx}
\par\smallskip
\small R: \path{raw_text}; S: summary; E: \path{embedding_text}; Vec: vector retrieval.
\end{table*}

\subsection{Statistical Reliability}
\begin{figure}[!t]
\centering
\includegraphics[width=0.98\columnwidth]{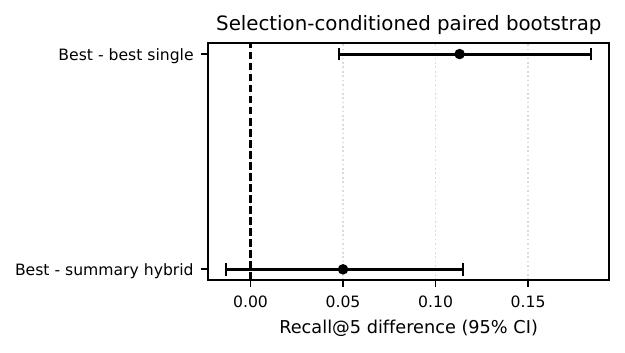}
\caption{Selection-conditioned paired-bootstrap 95\% confidence intervals for Recall@5 differences. Configuration-selection uncertainty is not included.}\label{fig:paired-diff}
\end{figure}

We fixed the best hybrid selected on the same 100 questions and ran 5,000 paired-bootstrap iterations. Its Recall@5 difference from \path{embedding_text_bm25} was +0.113, with a 95\% CI of [0.048, 0.184]. The difference from the summary hybrid at $\beta=0.50$ was +0.050, with a 95\% CI of [$-0.013$, 0.115] (Table~\ref{tab:bootstrap-diff}, Figure~\ref{fig:paired-diff}). Although the former interval excludes 0, it is a selection-conditioned interval that does not reflect the search over 126 configurations or multiple comparisons and is therefore not confirmatory evidence of superiority on unseen questions. No clear difference can be concluded for the latter comparison either.

\begin{table*}[!t]
\centering
\caption{Bootstrap difference confidence intervals}\label{tab:bootstrap-diff}
\small
\renewcommand{\arraystretch}{1.08}
\begin{tabularx}{\textwidth}{@{}lccX@{}}
\toprule
Comparison & $\Delta$Recall@5 & 95\% CI & Interpretation \\
\midrule
Best $-$ E-BM25 & +0.113 & [0.048, 0.184] & Positive conditional difference; selection uncertainty excluded \\
Best $-$ S-hybrid ($\beta=0.50$) & +0.050 & [$-0.013$, 0.115] & CI includes 0; inconclusive \\
\bottomrule
\end{tabularx}
\end{table*}

\subsection{Results by Question Type}
\begin{figure}[!t]
\centering
\includegraphics[width=0.98\columnwidth]{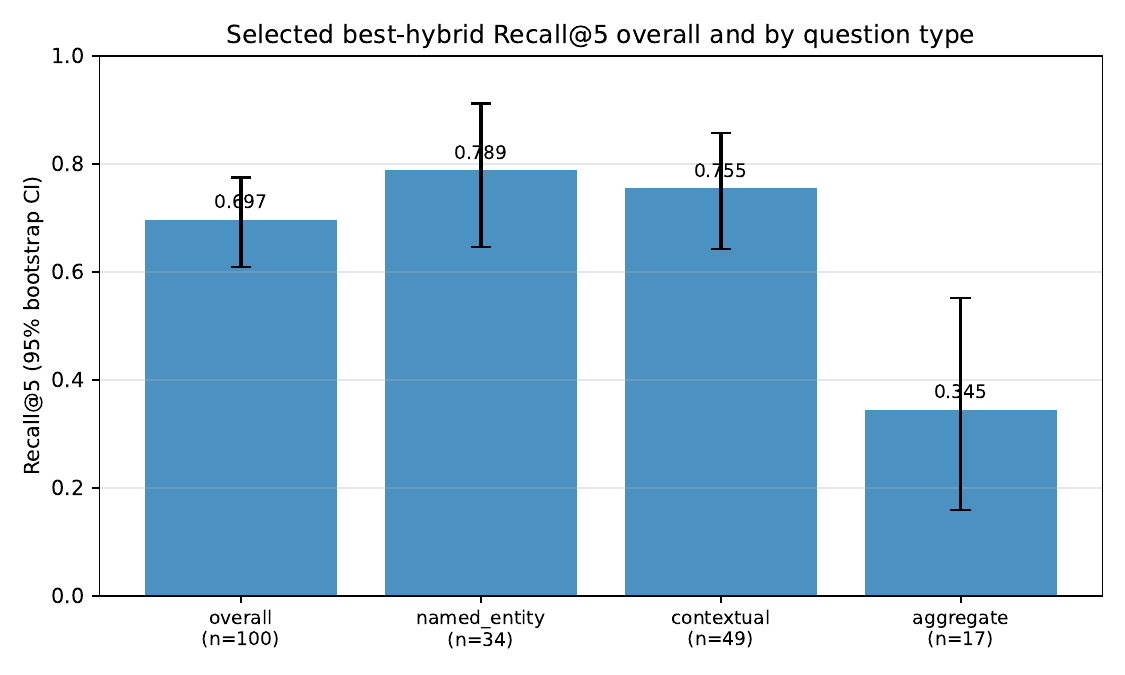}
\caption{Selection-conditioned Recall@5 for the selected best hybrid, overall and by question type.}\label{fig:question-type}
\end{figure}

Table~\ref{tab:question-type} and Figure~\ref{fig:question-type} show performance by question type. Recall@5 for the selected best hybrid was 0.789 [0.647, 0.912] on \path{named_entity}, 0.755 [0.643, 0.857] on contextual, and 0.345 [0.159, 0.551] on aggregate questions. Although there were only 17 aggregate questions and the confidence interval was wide, this case study yielded a lower point estimate for aggregate than for \path{named_entity} and contextual questions. We conducted no confirmatory test across question types, and these intervals are also conditional on the selected configuration.

\begin{table}[!t]
\centering
\caption{Selection-conditioned best-hybrid performance by question type}\label{tab:question-type}
\small
\renewcommand{\arraystretch}{1.08}
\begin{tabularx}{\columnwidth}{@{}lcc@{}}
\toprule
Question type & Recall@5 & 95\% CI \\
\midrule
\texttt{named\_entity} & 0.789 & [0.647, 0.912] \\
\texttt{contextual} & 0.755 & [0.643, 0.857] \\
\texttt{aggregate} & 0.345 & [0.159, 0.551] \\
\bottomrule
\end{tabularx}
\end{table}

\section{Failure Analysis}
\subsection{Top-5 Failures of the Selected Best Hybrid}
The selected best hybrid combines \path{embedding_text_bm25} and \path{embedding_text_vector} at $\beta=0.45$. For 22 of the 100 questions, its Top-5 results contained no primary-gold chunk.

By question type, failure occurred for 6/34 \path{named_entity}, 8/49 contextual, and 8/17 aggregate questions, giving aggregate the highest failure rate (Table~\ref{tab:best-failures}).

\begin{table}[!t]
\centering
\caption{Top-5 failures by question type under the selected best hybrid}\label{tab:best-failures}
\small
\renewcommand{\arraystretch}{1.08}
\begin{tabularx}{\columnwidth}{@{}Xrrr@{}}
\toprule
Question type & Questions & \shortstack{Top-5\\failures} & \shortstack{Failure\\rate} \\
\midrule
\texttt{named\_entity} & 34 & 6 & 17.6\% \\
\texttt{contextual} & 49 & 8 & 16.3\% \\
\texttt{aggregate} & 17 & 8 & 47.1\% \\
Overall & 100 & 22 & 22.0\% \\
\bottomrule
\end{tabularx}
\end{table}

\subsection{Aggregate Question Analysis}
Recall@5 for the selected best hybrid was 0.789 on \path{named_entity}, 0.755 on contextual, and 0.345 on aggregate questions. Eight of the 17 aggregate questions had no primary gold in the Top 5. This observation is specific to the present case study and does not imply generalization to other users or datasets.

Aggregate questions ask about personal tendencies, roles, or sustained activities, so their evidence is often distributed across chunks, dates, people, and conversation sources. Flat chunk-level retrieval may recover locally relevant chunks while still failing to place the multiple pieces of evidence needed for an answer near the top of the ranking.

This result suggests that hierarchical summaries, person- or project-specific timelines, query decomposition, and multi-hop retrieval may be useful for aggregate questions. None of these methods is evaluated here.

\subsection{Auxiliary Failure Analysis with the Summary Hybrid}
For a more detailed diagnostic labeling and Top-$K$ sensitivity analysis, we use the summary hybrid that combines \path{summary_bm25} and \path{summary_vector} at $\beta=0.50$.

This summary hybrid differs from the selected best hybrid in Section~6. The following results are auxiliary diagnostics for the summary hybrid, which failed at Top 5 on 30/100 questions. A script assigned mutually exclusive diagnostic labels in the following order: aggregate questions were labeled \path{aggregate_failure}; non-aggregate questions with at least three primary-gold chunks were labeled \path{multiple_primary_gold_nonaggregate}; and all remaining questions were labeled \path{other_rule_based_failure}. The respective counts were 9, 3, and 18 (Table~\ref{tab:failure-types}). \path{other_rule_based_failure} is a residual category that includes questions for which neither component retriever retrieved gold within the inspected range. These labels were assigned mechanically by rules and were not human-verified causal analyses; their names should not be interpreted as establishing the cause of an individual failure.

\begin{table}[!t]
\centering
\caption{Rule-based labels for 30 Top-5 failures ($\beta=0.50$)}\label{tab:failure-types}
\small
\renewcommand{\arraystretch}{1.08}
\begin{tabularx}{\columnwidth}{@{}Xrr@{}}
\toprule
Failure type & Count & Percentage \\
\midrule
\path{other_rule_based_failure} & 18 & 60.0\% \\
\path{aggregate_failure} & 9 & 30.0\% \\
\path{multiple_primary_gold_nonaggregate} & 3 & 10.0\% \\
\bottomrule
\end{tabularx}
\end{table}

The same analysis included 17 aggregate questions, with a mean of 2.35 primary-gold chunks. Nine spanned multiple dates and five spanned multiple sources. Of the nine aggregate questions that failed at Top 5, two were recovered by extending the cutoff to Top 10, and four in total were recovered by extending it to Top 20.

Increasing the number of retrieved chunks can therefore recover some aggregate questions, but questions that remain unrecovered at Top 20 may require redesign of the retrieval unit or memory structure rather than a simple increase in $K$.

\begin{figure}[!t]
\centering
\includegraphics[width=0.98\columnwidth]{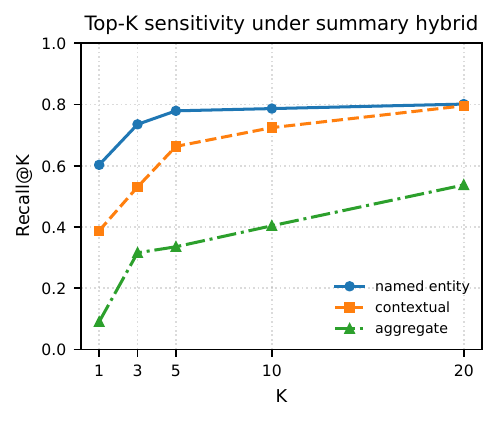}
\caption{Top-$K$ sensitivity by question type under the summary hybrid ($\beta=0.50$).}\label{fig:topk}
\end{figure}

\subsection{Retriever Characteristics}
Among the individual retrievers, \path{embedding_text_bm25} achieved Recall@5=0.584 and \path{embedding_text_vector} achieved 0.549, whereas their linear hybrid attained the highest point estimate of 0.697. Comparing Top-5 primary-gold hits by question, BM25 alone succeeded on 20 questions, vector retrieval alone on 16, both on 45, and neither on 19 (Table~\ref{tab:overlap}).

\begin{table}[!t]
\centering
\caption{Top-5 primary-gold success overlap for the selected retriever pair}\label{tab:overlap}
\begin{tabular}{lr}
\toprule
Success pattern & Questions \\
\midrule
BM25 only & 20 \\
Vector only & 16 \\
Both & 45 \\
Neither & 19 \\
\bottomrule
\end{tabular}
\end{table}

This auxiliary count shows that the retrievers succeed on different questions and is consistent with complementary retrieval patterns. It does not, however, causally attribute the hybrid improvement to complementarity alone: min--max normalization, candidate union, and changes in score distributions also act simultaneously.

\section{Discussion}
The high point estimates of configurations using \path{embedding_text} are consistent with the hypothesis that colocating a summary and a raw excerpt is useful. The summary describes the meaning and purpose of a conversation in natural language, whereas the raw excerpt preserves proper names, dates, event names, and colloquial expressions. However, because the fixed prefix, summary, and raw excerpt were not ablated separately, performance differences cannot be causally attributed to any one component.

While \path{raw_text} preserves observed vocabulary, it also contains short backchannels, omissions, typographical errors, and context-dependent expressions. A summary organizes meaning but may omit concrete terms. \path{embedding_text} places both together, while simultaneously changing document length and introducing a fixed prefix. The experiment therefore compares complete search representations rather than isolating the effect of each component.

Although \path{embedding_text_bm25} had the highest point estimate among individual retrievers, the difference between \path{embedding_text_vector} and \path{summary_vector} was small. The results suggest that the interaction between search representation and retriever may matter, but they do not establish that \path{embedding_text} is consistently superior across retrieval methods.

Future work should conduct a component ablation that compares a fixed prefix alone, summary alone, raw excerpt alone, and summary plus raw excerpt under the same retriever.

\subsection{Implications of Hybrid Retrieval}
The best hybrid explored on the same evaluation set used $\beta=0.45$, and its Recall@5 remained 0.687 at $\beta=0.50$. In this dataset, a linear combination that did not place nearly all weight on one component produced high point estimates. However, $\beta$ was selected in-sample, so the same weight is not necessarily optimal for unseen questions.

Combining lexical matching with semantic similarity is an explanatory hypothesis for the hybrid improvement, and the auxiliary Top-5 success-set counts are consistent with it. We did not conduct component ablations or a preregistered test of complementarity, so the source of the improvement cannot be uniquely identified.

In the bootstrap that fixed the selected best hybrid, the confidence interval for its difference from \path{embedding_text_bm25} excluded 0. This remains a selection-conditioned exploratory result and is not treated as confirmatory evidence of superiority. The confidence interval for the difference from the summary-based hybrid included 0, so that comparison was also inconclusive.

\subsection{Toward Personal Memory RAG}
LINE histories contain many short utterances, omissions, colloquial expressions, and context-dependent phrases. In this dataset, configurations using processed search representations attained higher point estimates than a configuration that searched only \path{raw_text} directly. This is nevertheless an exploratory single-user case study and does not establish a general advantage.

Among individual retrievers, \path{embedding_text_bm25} had the highest point estimate; among the explored hybrids, \path{embedding_text_bm25} + \path{embedding_text_vector} did so. This is consistent with the hypothesis that colocating a summary and raw excerpt and combining lexical and semantic matching is useful. However, the absence of component ablations and the small difference between \path{embedding_text_vector} and \path{summary_vector} leave the contribution of each element unresolved.

Aggregate questions performed worse than \path{named_entity} and contextual questions in this case study. When evidence is distributed across conversations, dates, people, and chunks, flat chunk retrieval with a single query may be insufficient.

Potential improvements include LLM-based query expansion, query rewriting, and query decomposition. Query2doc expands a query with an LLM-generated pseudo-document and reports improvements for both sparse and dense retrieval\cite{ref16}. Decomposing a broad, abstract question into multiple concrete search queries may likewise recover evidence distributed across chunks. For example, a question about a user's role in an activity could be decomposed into scheduling, requests, hands-on work, consultation, and reporting of outcomes. Such multi-query retrieval and query decomposition are candidates for improving aggregate questions, but they are not evaluated in this study.

Second-stage reranking of retrieved chunks and multi-hop retrieval that progressively retrieves and integrates multiple chunks also warrant investigation. Personal Memory RAG must represent not only isolated utterances but also developments over time, interpersonal relationships, and project progress. Hierarchical representations such as topic-level memory, person timelines, and project timelines may therefore complement chunk-level retrieval. This direction is shared by existing memory architectures for long conversations\cite{ref10,ref19}.

This retrieval-only study does not evaluate the accuracy, faithfulness to evidence, or usefulness of final LLM-generated answers. Future work should determine whether an LLM can generate accurate, evidence-attributed answers from the retrieved chunks. A personal-memory database must also be updated continuously as new conversations and activities occur. Integrating long-term memory updates, query generation, reranking, hierarchical memory representations, and citation-grounded answer generation is a next step toward Personal Memory RAG.

\section{Limitations and Ethical Considerations}
\subsection{Limitations}
This is an exploratory case study based on the LINE history of one user. Its results cannot be generalized to other users, social platforms, or languages; conversation partners, writing style, vocabulary, and activities vary across users.

NotebookLM was used to construct candidate evaluation questions from a JSON export containing all messages and their message IDs. The resulting question distribution may be biased toward formats that NotebookLM generates readily or toward salient topics and may not represent questions asked in actual use.

The gold labels were assigned by one annotator over a pool of 910 candidates, and all chunks outside that pool were treated as relevance 0. We did not verify exhaustive coverage of all relevant chunks or measure inter-annotator agreement. The same evidence appearing in overlapping windows was counted as separate chunks rather than consolidated into an evidence group, so duplication may affect the Recall denominator. In particular, evidence for aggregate questions is distributed, and the gold set may be incomplete.

Hybrid pairings and $\beta$ were explored on the same 100 questions. The reported best-hybrid performance and difference confidence intervals are conditional on the selected configuration and do not include the winner's curse or selection uncertainty introduced by the search. Confirmation on an independent held-out question set is required.

The BM25 system is a regex-tokenizer baseline without morphological analysis or character $n$-grams; it discards most Hiragana and single-character terms. This may disadvantage colloquial \path{raw_text}, so differences from summary-based retrieval cannot be attributed to representation alone. We also conducted no pre-specified audit of summary hallucinations or omissions.

Moreover, \path{embedding_text} is not a raw-free representation. It combines a fixed description, a summary, and an excerpt containing up to the first 4,000 Japanese characters of \path{raw_text}; it therefore retains raw-derived vocabulary and message content. No formal privacy protection is guaranteed.

The study is further limited to retrieval and does not evaluate the accuracy, evidence faithfulness, or utility of the final answers produced when retrieved chunks are supplied to an LLM.

Top-10/20 recoverability is a diagnostic analysis that widens the cutoff of the same summary-hybrid ranking; it is not an experiment with a different candidate pool. In addition, the run record does not contain the model snapshot used for summary generation, API retry history, the final input length used when the embedding cache was generated, or the SDK version at generation time. Dense retrieval also has no explicit tie-break for exactly equal scores. These omissions limit full reproducibility.

\subsection{Ethics and Privacy}
The data contain private communications involving both the researcher and third parties. To generate candidate evaluation questions, a JSON export of all LINE messages, each with a message ID, was supplied to NotebookLM. For summary generation, up to the first 7,000 Japanese characters of \path{raw_text} were sent to the OpenAI Responses API. For embedding generation, search representations containing a summary or raw excerpt were sent to the OpenAI Embeddings API. The study was therefore not conducted entirely locally and provides no formal privacy guarantee. Retrieval ranking and metric computation were performed using a local database and precomputed caches.

Explicit consent was not obtained from conversation partners for research use or for submission of their messages to NotebookLM and the OpenAI APIs. The study also did not undergo institutional ethics review. Although it was conducted as an exploratory self-case study by the researcher, these are substantial ethical limitations and do not establish the ethical acceptability of research involving third-party communications.

We do not release or redistribute the raw data, question text, gold annotations, raw previews, real names, LINE display names, source-file names, exact timestamps, or screenshots. Conversation examples in this paper are synthetic. Public results are limited to aggregates that exclude individual utterance text, question text, gold text, real names, source names, and exact timestamps. These exclusions do not formally guarantee that re-identification is impossible. HTML and CSV files containing raw previews were stored on a local computer accessible only to the researcher. The study did not establish dedicated encryption, a retention period, or a verified deletion procedure, which is an additional operational limitation.

The \path{embedding_text} representation is not raw-free and contains up to the first 4,000 Japanese characters of the original text. This study provides no formal anonymization, differential privacy, or privacy guarantee with respect to external services. Future work should consider explicit consent, institutional ethics review or a formal determination that review is not required, data minimization, and local models before data processing begins.

Accordingly, the public materials for this paper contain no raw data or individual evaluation questions. They are limited to aggregate results, implementation specifications, synthetic examples, and question-ID-level metrics with question text and related content removed. Restricting the release in this way does not provide a formal privacy guarantee.

\section{Conclusion}
As an initial step toward Personal Memory RAG for LLMs, this study conducted a retrieval-only evaluation over a personal LINE conversation history. We segmented 358,896 messages from one user into 22,329 chunks and evaluated whether conversation chunks supporting answers could be retrieved for 100 questions.

We compared \path{raw_text}, summary, and \path{embedding_text} as search representations. \path{embedding_text_bm25} attained the highest point estimate among individual retrievers. The \path{embedding_text_bm25} + \path{embedding_text_vector} hybrid selected at $\beta=0.45$ on the same 100 questions achieved Recall@5=0.697, exceeding the best individual retriever by 0.113 with a selection-conditioned 95\% CI of [0.048, 0.184]. Because the interval excludes uncertainty from configuration and $\beta$ selection, it must be interpreted as exploratory.

The selected best hybrid had a higher point estimate than the summary hybrid, but the confidence interval for their difference included 0. The individual contributions of the summary, raw excerpt, and fixed prefix within \path{embedding_text} also remain unidentified because no component ablation was performed.

Aggregate questions showed lower performance in this case study. Topic-level memory, person and project timelines, multi-hop retrieval, and query decomposition are candidate improvements for future evaluation.

Future work should extend the retrieval foundation evaluated here into a complete Personal Memory RAG pipeline that supplies retrieved chunks to an LLM for answer generation. Integrating evidence-attributed answers, long-term memory updates, hierarchical memory representations, and LLM-based search-query generation may make personal conversation histories more useful as external memory.

\appendix
\section{Summary Generation Prompt}
For reproducibility, the prompt used for summary generation is described below. The original prompt was written and executed in Japanese; the following is an English translation, and the exact Japanese text is included in the source package as \path{summary_prompt_ja.txt}. The runtime requested \path{gpt-4o-mini}, set \path{max_output_tokens}=700, a timeout of 90 seconds, and a 0.2-second call interval, and did not set temperature, top-p, seed, or an explicit retry count. Per-chunk response-model snapshots were not stored.

\begin{quote}\small
\textbf{Instructions: }You are responsible for building a memory-search index for a personal AI. Read a chunk of original LINE messages and write a natural-language summary in Japanese that will be suitable for later semantic search. Preserve proper names, interpersonal relationships, topics, emotions, commitments, events, and contexts such as romance, school, and work. Do not speculate excessively; include only information supported by the original text.\par\medskip
\textbf{User input template: }Summarize the following chunk of original LINE messages as natural-language text for retrieval. Write approximately 200--500 Japanese characters. After a blank line, the program appended up to the first 7,000 Japanese characters of \path{raw_text}.
\end{quote}
\section{Experimental Environment and Provenance}
\begin{table}[H]
\centering
\caption{Audited execution environment}
\small
\begin{tabularx}{\columnwidth}{@{}p{0.30\columnwidth}X@{}}
\toprule
Item & Recorded value \\
\midrule
Operating system & Microsoft Windows 11 Home, build 26200 \\
Python & 3.12.13 \\
OpenAI Python SDK & 2.36.0 (audited environment; generation-time version not recorded) \\
NumPy & 2.4.6 \\
SQLAlchemy & 2.0.49 \\
Database & SQLite via SQLAlchemy \\
Repository commit & Not available; the research copy was not a Git worktree \\
\bottomrule
\end{tabularx}
\end{table}

The exact generation prompt supplied to NotebookLM and the provider-side model version were not saved in the experimental record. Candidate-question generation therefore cannot be reproduced exactly. For the final 100 questions, a supplementary question-ID-level metric CSV was saved with raw text and question text removed.
\FloatBarrier
\balance

\end{document}